\documentclass[aps,prl,twocolumn,showpacs,amsmath,floatfix]{revtex4}
\usepackage{graphicx}
\usepackage{dcolumn}

\def\ee{\boldsymbol{\cal{E}}}

\begin{document}
% \topmargin-0.8cm

% \draft

\title{
Predicting polarization and nonlinear dielectric response of
arbitrary perovskite superlattice sequences }

\author{Xifan Wu$^{1}$, Massimiliano Stengel$^{2}$,
Karin M. Rabe$^{3}$, and David Vanderbilt$^{3}$ }

\address{$^{1}$Chemistry Department, Princeton University, Princeton, NJ
08544-0001,USA}

\address{$^{2}$Materials Department, University of California,
Santa Barbara, CA 93106-5050,USA}

\address{$^{3}$Department of Physics and Astronomy, Rutgers University,
Piscataway, NJ 08854-8019, USA}

\date{\today}

\begin{abstract}

We carry out first-principles calculations of the nonlinear
dielectric response of short-period ferroelectric superlattices.
We compute and store not only the total polarization, but also
the Wannier-based polarizations of individual atomic layers, as a
function of electric displacement field, and use this information
to construct a model capable of predicting the nonlinear dielectric
response of an arbitrary superlattice sequence.
We demonstrate the successful application of our approach to
superlattices composed of SrTiO$_3$, CaTiO$_3$,
and BaTiO$_3$ layers.
%
% This first-principles modeling approach
% provides a powerful tool for the theoretical prediction and design
% of complex ferroelectric superlattice structures.
%
\end{abstract}

%\pacs{77.22.-d, 77.22.Ej, 77.80.-e, 77.84.Bw}
\pacs{77.22.-d, 77.22.Ej, 77.80.-e, 77.84.Lf}

\maketitle

%% \narrowtext

\marginparwidth 2.7in

\marginparsep 0.5in

\def\dvm#1{\marginpar{\small DV: #1}}
\def\kr#1{\marginpar{\small KR: #1}}
\def\msm#1{\marginpar{\small MS: #1}}
\def\xwm#1{\marginpar{\small XW: #1}}

% \def\dvm#1{}
% \def\kr#1{}
% \def\msm#1{}
% \def\xwm#1{}

%%%%%%%%%%%%%%%%%%%%%%%%%%%%%%%%%%%%%%%%%%%%%%%%%%%%%%%%%%%%

The development of advanced methods for layer-by-layer epitaxial
growth of multicomponent perovskite superlattice structures
has generated excitement \cite{Review_mod},
both because of the intriguing materials physics that comes into
play, and because of potential applications in non-volatile
ferroelectric memories, piezoactuators and sensors, and
magnetoelectric devices \cite{Scott}.  To guide experimental
exploration of this greatly expanded class of materials,
there is a critical need for atomic-scale understanding
and modeling of key structural and functional properties,
particularly polarization and dielectric response.

First-principles methods have allowed for the direct quantitative
computation of such material-specific information for
representative perovskite superlattices \cite{Jeff,Bousquet}.
However, such calculations are limited to relatively short-period
superlattices, of the order of ten unit cells.  First-principles
{\it modeling} can extend our theoretical capability so that one can
make predictions about arbitrary stacking sequences and elucidate
the physics behind the novel behavior of superlattices.  In
particular, substantial progress has recently been made in isolating
and studying the effects of the epitaxial strain on film structure,
polarization, and piezoelectric properties
\cite{Review_strain,Oswaldo_strain}.

It is clear, however, that it is electrostatic effects that dominate
the physics of superlattices built from ferroelectric
(e.g., BaTiO$_3$) and incipient ferroelectric (e.g., SrTiO$_3$)
constituents.  In previous first-principles models,
these effects were included in an approximate way, either by
describing the layers in terms of their bulk linear dielectric
properties \cite{Jeff}, or by
imposing a constant-polarization layer-to-layer constraint that only
roughly captures the effects of the internal electric fields
\cite{Serge}. Furthermore, most previous first-principles
calculations, using the periodic boundary conditions implicit in
ordinary implementations, give results only for zero applied
electric field.
Since much of the interest in perovskite superlattices lies
in their use in capacitor structures whose performance
relies on their nonlinear dielectric behavior under bias
voltage, a more fundamental methodology capable of capturing
such effects is urgently needed.

In this Letter, we present a rigorous first-principles treatment
allowing computation and modeling of the electrostatics and non-zero
electric-field response of perovskite oxide superlattices.
Our approach is based on a recently-developed
Wannier-based formulation of the layer polarizations in perovskite
superlattices \cite{Xifan_layer} in combination with methods for
treating insulators in finite electric fields
\cite{Souza,Umari,Constrain_P,Max_field}.
Crucially, we choose here to work at fixed electric {\it displacement}
field \cite{Max_fixedD}, and show that this gives
a clean separation between long-range Coulomb interactions and
short-range interfacial effects.  As we demonstrate through
application to superlattices composed of three ABO$_3$ perovskite
constituents, the resulting model yields, for arbitrary stacking
sequences, quantitative predictions of polarization and nonlinear
dielectric response with \emph{ab-initio} accuracy, thus enabling the
theory-driven search of the full range of superlattice sequences for
novel or optimized properties.

The construction of our model begins with the decomposition of the
superlattice into atomic layers, specifically into AO and BO$_2$
layers alternating along [001].  The individual layer polarizations
(LP) for each AO and BO$_2$ layer $j$ are computed using the
Wannier-based method
of Ref.~\cite{Xifan_layer}, and recorded as functions $p_j(D)$ of
the displacement field $D$ using a constrained-$D$ first-principles
implementation \cite{Max_fixedD}.  This choice is appropriate because
(i) $D$ is constant throughout the supercell
($\nabla\cdot{\bf D} = 4\pi\rho_{\rm free}=0$), while the local
macroscopic $\cal E$ and $P$ generally vary, and (ii)
imposing constant-$D$ electrical boundary conditions has the 
virtue of making the force-constant matrix of the quasi-one-dimensional 
superlattice short-ranged in real space.   This ``locality principle''
implies that one may expect the $p_j(D)$ to depend 
only on the \emph{local} compositional environment comprising the layer 
itself and few nearby neighbors.
For any given superlattice, the total polarization, which is a
quantity of central interest, is given by
$P(D)=c(D)^{-1}\sum_{j}p_j(D)$.
It is also straightforward
to obtain the electric equation of state in other forms, e.g.,
$D(P)$ by numerical inversion, and ${\cal E}(P)=D(P)-4\pi P$.

We demonstrate the method through application to superlattices
comprised of an arbitrary sequence of SrTiO$_3$, BaTiO$_3$ and
CaTiO$_3$ layers, grown in the (001) direction with an in-plane
lattice constant $a_0$ = 7.275 bohr, our theoretical equilibrium
lattice constant of bulk SrTiO$_3$.  We assume
1$\times$1 in-plane periodicity and tetragonal $P4mm$ symmetry, thus
neglecting possible intermixing, nanodomain formation, or the
appearance of antiferroelectric or octahedral-tilting distortions.
Some of these effects may be important in real superlattices and
will deserve attention in future extensions, but our focus in the
present work is to isolate and study the effects associated with the
complex stacking of ideal layers.  Consistent with the $P4mm$
symmetry, $D$ is taken along the $z$ axis, ranging in steps from
$-$0.32 to 0.32\,C/m$^2$.

First-principles calculations to optimize the structure for fixed 
$D$ \cite{Max_fixedD} and to obtain the layer polarizations $p_j(D)$
and lattice constants $c(D)$ were performed on a database of
superlattices using the {\tt Lautrec} code package, which
implements plane-wave calculations in the PAW framework
\cite{Bloechl_PAW} in the LDA approximation \cite{LDA_PW92}.
The polarization and its coupling to the electric field is handled
by an efficient real-space Wannier formulation~\cite{Max_field}.
We used a plane-wave energy cutoff of 40 Ry and a 6$\times$6$\times$2
Monkhorst-Pack $k$-mesh.
The database of superlattice structures contains all one- and
two-component period-4 supercells (BBBB, SSSS, CCCC, BBBS, BBSS,
BSBS, BSSS, CCCS, CCSS, CSCS, CSSS, BBBC, BBCC, BCBC, BCCC), and one
three-component superlattice SSBC, where C, S, and B refer to
CaTiO$_3$, SrTiO$_3$, and BaTiO$_3$ layers respectively.

\begin{figure}
\includegraphics[width=2.5in]{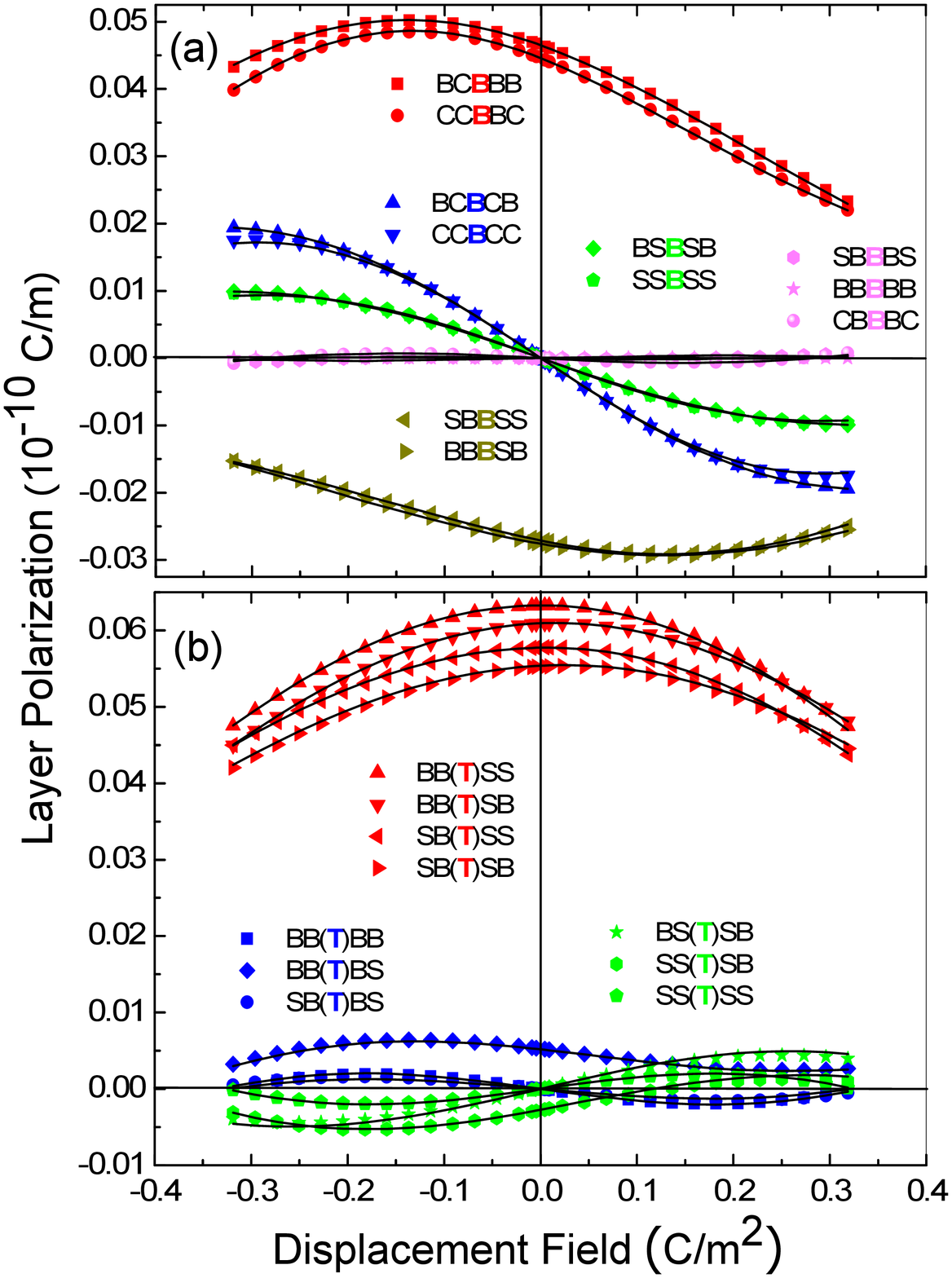}
\caption{\label{fig1} (Color online.)  Dependence of layer
polarizations on chemical environment for (a) BaO planes
(relative to a BaO plane in bulk BaTiO$_3$), and
(b) TiO$_2$ planes (relative to an average of TiO$_2$
planes in bulk BaTiO$_3$ and SrTiO$_3$).  C, S, B, and T
refer to CaO, SrO, BaO, and TiO$_2$ layers, respectively.
First-principles results and model fittings are denoted by symbols
and solid lines respectively.}
\end{figure}

Representative $p_j(D)$ curves are presented in
Fig.~\ref{fig1}.  It is striking that the LP curves separate,
as expected from our locality principle, into clusters depending on the
nearest-layer chemical environment (color-coded
for comparison).  However, the differences among the curves
within a cluster are still too large to neglect, especially for TiO$_2$
layers, indicating that an
accurate model must include further-neighbor interactions as well.
The effects of local inversion symmetry breaking
are also clearly visible.  For example, a BaO layer in
the middle of a CBB sequence has a large and positive LP
even at $D$=0.  Smaller shifts arise from the second-neighbor
environment, e.g., for the central TiO$_2$ layer in a
BBSS sequence.

We now introduce a cluster expansion (CE) \cite{CE_Zunger} for
the environment dependence of the $p_l(\{s\};D)$ as
\begin{eqnarray}
p_l (\{s\})& = & J_0 
   + \sum_{i}\left(  J_{l,i} s_i + J_{l,i}^{\prime}s_i^2 \right) \cr
  & + & \sum_{ij} \left( J_{l,ij} s_is_j
   +  J_{l,ij}^{\prime}s_is_j^2
   +  J_{l,ij}^{\prime\prime}s_i^2s_j^2 \right) \cr
  & + & \sum_{ijk}J_{l,ijk} s_is_js_k + ... ,
\label{eq:CE_general}
\end{eqnarray}
where the $J$ are $D$-dependent effective cluster interactions (ECI)
to be determined from fitting to the first-principles database.
We choose ``spin'' variables $s_i=-$1, 0, and 1 to identify AO layer
$i$ as  CaO, SrO, and BaO repsectively, to reflect the
fact that Sr is midway between Ca and Ba in the Periodic Table.
Thus, insofar as Sr acts like an average of Ca and Ba atoms, each
appearance of a squared spin variable $s^2$ in a term of the CE
makes it more likely that the term can be neglected.

We therefore approached the truncation and fitting of the model of
Eq.~(\ref{eq:CE_general}) with three principles in mind:
(1) the importance of an $n$-body term is expected to decrease
rapidly with $n$;
(2) the dependence of $p_l$ on $s_i$ should decay
rapidly with the distance between layers $l$ and $i$; and
(3) coefficients with prime superscripts, corresponding to ``higher-level''
spin variables $s^2$, should be less important than those without.

Translation and spatial-inversion symmetry imply that
$J_{l,l+m}(D) = -J_{l,l-m}(-D)$,
$J_{l,l+m,l+n}(D) = -J_{l,l-m,l-n}(-D)$, etc., independent of $l$.
It is therefore natural to define
${\cal J}_m=(J_{l,l-m}+J_{l,l+m})/2$ and
$\widetilde{\cal J}_m=(J_{l,l-m}-J_{l,l+m})/2$, and
similarly for two-body and higher terms. Correspondingly, we
separate Eq.~(\ref{eq:CE_general}) into parts
$p_{\rm AO}^{(-)}(D)$ and $p_{\rm TiO_2}^{(-)}(D)$
that are odd in $D$, and parts
$p_{\rm AO}^{(+)}(D)$ and $p_{\rm TiO_2}^{(+)}(D)$
that are even in $D$,
reflecting the inversion-symmetry-conserving 
and inversion-symmetry-breaking characters of the local environment
respectively.  To give a sense of the form of the resulting
expressions, the odd part for AO layers becomes
\begin{eqnarray}
p_{\rm AO}^{(-)} & = &
{\cal J}  +  {\cal J}_0s_0 + {\cal J}_0's_0^{\,2} + {\cal J}_{1}(s_{\bar{1}} + s_1)
+  {\cal J}_{1}'(s_{\bar{1}}^{\,2} + s_1^{\,2}) \cr
&+& {\cal J}_2(s_{\bar{2}} + s_2) + {\cal J}_{01}(s_{\bar{1}}s_0+s_0s_1) \cr
&+& {\cal J}_{02}(s_{\bar{2}}s_0 + s_0s_2)+ {\cal J}_{\bar{1}1}s_{\bar{1}}s_1 \cr
&+& {\cal J}_{12}(s_{\bar{2}}s_{\bar{1}} + s_1s_2)
\label{eq:AO_odd}
\end{eqnarray}
where $\bar{n}$=$-n$ and layer 0 is the one whose LP is being expanded.
Similar expressions for $p_{\rm AO}^{(+)}$, $p_{\rm TiO_2}^{(-)}$,
and $p_{\rm TiO_2}^{(+)}$ are given in the supplementary material
\cite{EPAPS}.
The supercell lattice constant $c(D)$ is correspondingly expanded as
\begin{equation}
c=\sum_j({\cal C}_1+{\cal C}_2s_j+{\cal C}_3s_j^2+{\cal C}_4s_js_{j+1}) \,,
\label{eq:C_CE}
\end{equation}
where ${\cal C}_1(D)$, ${\cal C}_2(D)$, and ${\cal C}_3(D)$
assign to each layer its bulk $c(D)$, and only the
${\cal C}_4(D)$ term includes true superlattice effects.
The ${\cal J}(D)$ and ${\cal C}(D)$ parameters are expressed as
fifth and fourth-order Taylor expansions in $D$ respectively,
with the Taylor coefficients obtained
by fitting to the first-principles calculations of $p_j(D)$ and
$c(D)$ for superlattices in the database.

The choice of terms
to include in Eq.~(\ref{eq:AO_odd}) and in the corresponding
expressions for $p_{\rm AO}^{(+)}$, $p_{\rm TiO_2}^{(-)}$,
and $p_{\rm TiO_2}^{(+)}$
\cite{EPAPS} have been obtained using linear regression
techniques to identify higher terms that could be deleted
without significantly degrading the quality of the fit.
The linear-in-$D$ coefficients of the ECIs in
Eq.~(\ref{eq:AO_odd}) are presented in Table \ref{table:ECI},
confirming our expectation that terms of higher body, longer
range, and higher level tend to be less important.  Tables
listing values of all of the ECI coefficients are provided in the 
supplementary material \cite{EPAPS}.

The quality of the fit is excellent; the overall RMS error
in $p_j(D)$ values relative to first-principles results
is $2 \times 10^{-14}$\,C/m
for structures in the database.  This is illustrated
in Fig.~\ref{fig1}, where the solid lines representing the
fitted functions can be seen to pass quite accurately through
the first-principles symbols.  The quality of the fit is
similar for other cases, not plotted.

\begin{table}
\caption{Fitted linear-in-$D$ term of effective cluster
interactions for AO layers as defined in Eq.~\ref{eq:AO_odd}.}
\begin{ruledtabular}
\begin{tabular}{lcdcd}
& ECI & \multicolumn{1}{c}{\quad Value} & ECI &
        \multicolumn{1}{c}{\quad Value} \\
\hline
Zero-body
& $\cal J$      & 2.2771   &   \\
One-body
& ${\cal J}_0$    & 0.1113   & ${\cal J}_0'$    & 0.0819 \\
& ${\cal J}_1 $   & 0.0034   & ${\cal J}_1'$   &  0.0007 \\
& ${\cal J}_2$    & -0.0018  &          &          \\
Two-body
& ${\cal J}_{01}$ & 0.0197   & ${\cal J}_{02}$    & 0.0031 \\
& ${\cal J}_{\bar{1}1}$ & 0.0026 & ${\cal J}_{12}$ & 0.0013 \\
\end{tabular}
\end{ruledtabular}
\label{table:ECI}
\end{table}

\begin{figure}
\includegraphics[width=2.8in]{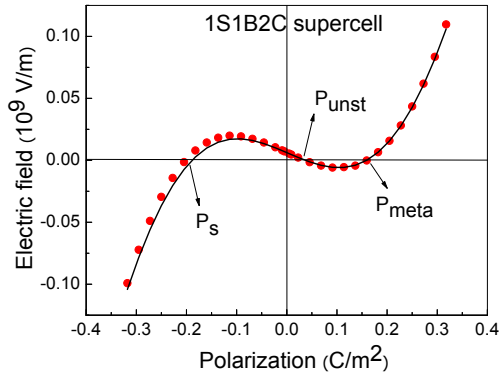}
\caption{\label{fig2} (Color online.) Model prediction (solid
lines) and first-principles calculations (symbols) of $\ee$ vs.~$P$
for 1S1B2C superlattice.}
\end{figure}

The model can now be used to predict the nonlinear dielectric
and piezoelectric properties of arbitrary superlattice sequences.
To illustrate the quality of the fit for supercell configurations
that were {\it not} included in the fit, we compare
in Fig.~\ref{fig2} the first-principles $P({\cal E})$ curves with
the model fits for the tri-color 1S1B2C supercell.  The
$P(\ee)$ curves are seen to be in excellent agreement.
The arrows in Fig.~\ref{fig2} indicate ${\cal E}$=0 solutions
corresponding to $P$=$-P_{\rm s}$ in the preferred (stable)
phase, $P$=$P_{\rm unst}$ at the unstable saddle point, and
$P$=$P_{\rm meta}$ in the metastable phase.
Note that $|P_{\rm s}|\ne|P_{\rm meta}|$ and $P_{\rm unst}\ne 0$
because the superlattice has broken inversion symmetry.
The model predicts $P_{\rm s}$, $P_{\rm unst}$, and $P_{\rm meta}$
values of $-$0.19, 0.04, and 0.16\,C/m$^2$, to be compared with direct
first-principles values of $-$0.20, 0.04, and 0.16\,C/m$^2$, respectively.
While the inversion-symmetry-breaking effects are subtle,
they are critical for tuning certain ferroelectric properties
\cite{Lee_nature,Sai},
and it is gratifying that they are obtained
accurately by our model.

\begin{table}[b]
\caption{Predicted magnitude of polarization (C/m$^2$)
for preferred ($P_{\rm s}$) and metastable ($P_{\rm meta}$)
polarization states in $n$S$n$B$n$C superlattices.}
\begin{ruledtabular}
\begin{tabular}{lrrrrr}
 & $n$=1\; & $n$=2\; & $n$=4\; & $n$=8\; & $n$=$\infty$\; \\
\hline
$P_{\rm s}$    &  $-$0.040 &    $-$0.198 &    $-$0.250 &    $-$0.267 &    $-$0.274 \\
$P_{\rm meta}$ &        &  0.178 & 0.237 & 0.262 & 0.274 \\
\end{tabular}
\end{ruledtabular}
\label{table:ps}
\end{table}

\begin{figure}[t]
\includegraphics[width=3.4in]{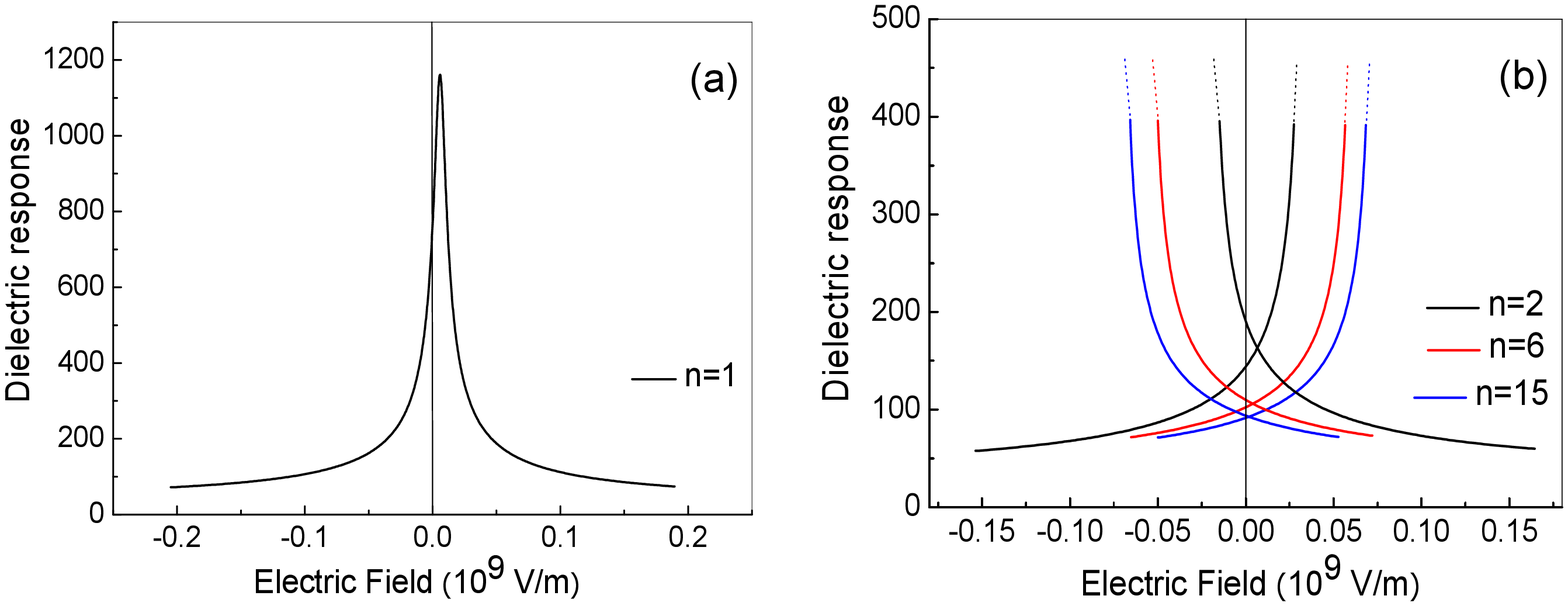}
\caption{\label{fig3} (Color online.)  Dielectric response
of $n$S$n$B$n$C superlattice in (a) paraelectric and
(b) ferroelectric regime.}
\end{figure}

To demonstrate the ability of our model to predict properties of
long-period superlattices that would be impractical for
direct first-principles calculations, we present our model predictions
for the spontaneous polarizations and dielectric responses
(including the piezoelectrically mediated component)
of $n$S$n$B$n$C superlattices in Table~\ref{table:ps} and
Fig.~\ref{fig3} respectively.
Because of the broken inversion symmetry, one polarization
direction is favored, with a different magnitude, over the
other. The polarizations approach the bulk value of 0.274\,C/m$^2$
for large $n$, but with decreasing $n$ we find a progressive
suppression of the polarization and an enhancement of
the asymmetry until, at $n$=1, the system becomes paraelectric,
with a single minimum.
These effects are also evident in the
dielectric response curves shown in Fig.~\ref{fig3},
where it is clear that the curves lack
reflection symmetry about the vertical axis, and the system
is seen to cross from ferroelectric to paraelectric behavior
between $n$=2 and $n$=1.

To gain more insight into interface effects in these superlattices,
we can characterize each of the six kinds of interfaces by an interface
dipole, extracted from our first-principles model as follows.
Using S/B as an example, we imagine an infinite stack ...SSSBBB...
with one interface, and define
\begin{equation}
p_{\rm int}(D)=\sum_i p_i(D) - p^{(0)}_{i}(D)
\label{eq:dipole_inter}
\end{equation}
where $i$ runs over AO and TiO$_2$ layers, $p_i(D)$ is the actual
layer polarization predicted by our model, and $p^{(0)}_{i}(D)$
is the polarization of that layer type in its own bulk environment.
(For the central TiO2 layer, $p^{(0)}(D)$ is the average of the
bulk S and B values.)  Because of the short-range nature of the
model, the sum only needs to run over a few layers near
the interface.

\begin{figure}[ht]
\includegraphics[width= 2.8in]{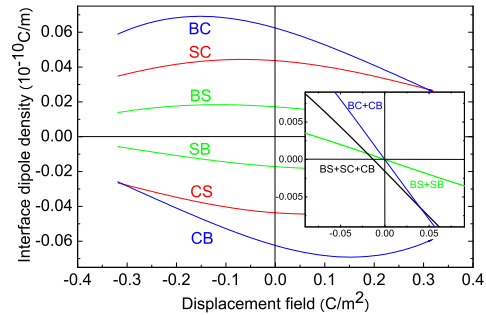}
\caption{\label{fig4}(Color online.) Model interface dipole densities.}
\end{figure}

The resulting $p_{\rm int}(D)$ curves are
presented in Fig.~\ref{fig4}.  First, note that the curves tend
to have a negative slope at $D$=0; this reflects the fact that the
presence of interfaces tends to suppress the ferroelectricity,
as was also evident in Table~\ref{table:ps} and Fig.~\ref{fig3}.
Second, each pair such as BS and SB are related by the symmetry
$p^{\rm BS}_{\rm int}(D)=-p^{\rm SB}_{\rm int}(-D)$
required by the condition that the overall $P(D)$ of a bicolor
$n$S$n$B superlattice must be odd in $D$.  Third, the inversion
symmetry breaking, relevant to tricolor superlattices such as
$n$B$n$S$n$C, is evident in the failure of the
$p^{\rm BS}_{\rm int}(D)+ p^{\rm SC}_{\rm int}(D)+
p^{\rm CB}_{\rm int}(D)$ curve to pass through the origin in
the inset of Fig.~\ref{fig4}.

The utility of the interface dipole concept is that, for any
superlattice in which the interfaces are never separated by less
than three unit cells (as determined by the range of our model),
$P(D)$ can be calculated just
by summing the bulk contribution for each layer and then adding
the contribution from each interface.  This prescription yields
a simplified but rigorously equivalent model that can be used in
such cases.  Thus, for example,
the inversion-symmetry-breaking effects in $l$B$m$S$n$C
superlattices are captured exactly by the simplified prescription
as long as $l$, $m$, $n\ge3$.

In summary, we have shown how a model can be extracted from
first-principles calculations on short-period superlattices and
used to make quantitatively accurate predictions of nonlinear
dielectric and piezoelectric responses over a range of applied
fields for arbitrary superlattice sequences.  The treatment of
electrostatic effects is rigorous, a key aspect being the choice
of the displacement field as the fundamental electrical variable
so as to keep interlayer interactions short-ranged.  The approach
can be straightforwardly generalized to include dependence on
epitaxial strain.  Such an approach can play an important role in
enabling the design of multifunctional ferroelectric superlattices
with desired polarization, piezoelectric or dielectric responses.

\acknowledgments
This work was supported by ONR grants N00014-05-1-0054 and
N0014-00-1-0261, and ARO-MURI grant W91NF-07-1-0410.

%%%%%%%%%%%%%%%%%%%%%%%%%%%%%%%%%%%%%%%%%%%%%%%%%%%%%%%%%%%%

\end{document}